\documentclass[10pt,a4paper,twoside]{article}
\usepackage{epsfig}
\usepackage{baltlat5}
\usepackage{wrapfig}
\begin{document}
\ \
\vspace{-0.5mm}

\setcounter{page}{601}
\vspace{-2mm}

\titlehead{Baltic Astronomy, vol.\ts 15, 601--604, 2006.}

\titleb{RADIATIVE TRANSFER PROBLEM IN DUSTY GALAXIES:\\
A PROGRAM SUITE}

\begin{authorl}
\authorb{Dmitrij~Semionov}{} and
\authorb{Vladas~Vansevi{\v c}ius}{}
\end{authorl}

\moveright-3.2mm
\vbox{
\begin{addressl}
\addressb{}{Institute of Physics, Savanori{\c u} 231, Vilnius LT-02300, Lithuania \\ dima@astro.lt}
\end{addressl}
}

\submitb{Received 2006 November 29; accepted 2006 December 10}

\begin{summary}
We present a program suite for the radiative transfer problem solution
in axisymmetrical dusty galaxy disks, intended primarily for
spectrophotometric analysis of stellar populations by means of
integrated and differential photometry. The solution is obtained
using a 2-D ray-tracing algorithm at a discrete wavelength set,
emphasizing careful treatment of the effects of light scattering
by interstellar dust grains. The program has been thoroughly tested
and shows the performance and accuracy comparable to or better
than other codes currently in use for astrophysical radiative transfer.
The program's source code and example model files are available at
{\tt http://www.astro.lt/\~{}dima/gfe/}
\end{summary}


\resthead{Radiative transfer problem in dusty
galaxies}{D.~Semionov, V.~Vansevi{\v c}ius}

\sectionb{1}{INTRODUCTION}

In order to correctly interpret the observed spectrophotometric
properties of galaxies and thus understand their evolution it is
necessary to investigate the inter-relations between  star formation
histories and  galaxy color/spectral evolution.  However, it is
impossible to establish these relations solely by modeling the colors of
stellar populations.  Interstellar extinction -- especially, the
scattering of stellar light by interstellar grains -- plays an important
role in redistributing the radiative energy and determining the radial
color gradients in galactic disks.  The only way to account for the
effects of the galaxy internal extinction is a thorough and rigorous
solution of the radiative transfer problem in dusty environments.

For efficient analysis of star formation history in galaxies it is
important to obtain self-consistent solutions of the radiative transfer
problem, i.e., the resulting derived spectral properties of the model
galaxies must depend only on the input spectra of stellar populations,
their spatial distribution and assumed properties of the interstellar
dust.  The code for the radiative transfer problem solution has to be as
efficient as possible since, for an investigation of a single object or
an analysis of a galaxy survey, we must account for a large number of
free parameters:  stellar population and dust distribution scale-lengths
and scale-heights, the amount of dust, extinction laws, as well as
various evolutionary parameters of stellar populations.

Most of the currently used astrophysical radiative transfer codes
employ either the Monte-Carlo (e.g., Ciardi \etal\ 2001) or the
ray-tracing (e.g., Razoumov \& Scott 1999) methods. Some of these
codes were compared and their differences discussed by Ivezic 
\etal\ (1997) and Baes \& Dejonghe (2001) for 1-D and by Dullemond
\& Turolla (2000) and Pascucci \etal\ (2004) for 2-D cases.

The Galactic Fog Engine (hereafter GFE), a program for a self-consistent
solution of the radiative transfer problem in axisymmetrical models of
dusty galaxy discs, presented in this paper, was developed considering
all the requirements of careful treatment of the radiative transfer
problem, mentioned above.  The details of the ray-tracing scheme,
implemented in GFE, were described by Semionov \& Vansevi{\v c}ius
(2005c).  Here we describe the GFE from the point of view of the
prospective user, outlining the steps required to construct galaxy
models for evaluation of spectrophotometric effects of internal
extinction by diffuse interstellar dust.

\sectionb{2}{THE GALACTIC FOG ENGINE}

In order to correctly recover histories of star formation in galaxies a
wide range of spectral and detailed spatial energy distribution
parameters are required.  A far-infrared flux of the galaxy is
particularly important to estimate its dust content.  However, this
information is not always available from observations and, furthermore,
to interpret it correctly a detailed knowledge of the dust composition
and its emissivity properties is required.  Therefore, we restricted
consideration of the galaxy models into UV, visual and NIR spectral
ranges, which are dominated by stellar light -- either directly observed
or scattered by interstellar dust grains.  The deterministic nature and
a possibility of effective optimization, inherent in the ray-tracing
approach, were the reasons to employ this algorithm for computation of
the scattered light distribution using an iterative scheme (Henyey
1937).

The model galaxy is represented by a cylinder, subdivided into a
set of layers of concentric internally homogeneous rings of
varying vertical and radial extent. The amount and distribution of
the absorbed and scattered light is obtained by sampling light
and dust distributions within a cylinder, using a set of rays along
which 1-D radiative transfer problems are solved. This ray-tracing
iteration is repeated for a desired number of times by substituting
the scattered light for the radiative field distribution. The GFE source
code, helpful model data managing utilities, a user's manual and model
examples are available at the project's web page,
{\tt http://www.astro.lt/\~{}dima/gfe/}

GFE results were compared in detail with other radiative transfer
problem solving code outputs by reproducing several previously published
1-D and 2-D models of dusty environments (Semionov \& Vansevi{\v c}ius
2002; an update to this article is available at the project's web
page). The results, obtained using the GFE, provide a very close match
(within 1\%) to the results, obtained using Monte-Carlo method
(Semionov \& Vansevi{\v c}ius 2002, 2005a,b).

The program is written in Fortran~77 and has been thoroughly tested on
various desktop and workstation configurations using GNU g77 and other
free and proprietary compilers with consistent results and performance.
The code is free for science and academic use, however, publications,
using the data computed by GFE, must acknowledge the code's use by
including a reference to the present article.  Also, at the first
mention of the Galaxy Fog Engine or the GFE abbreviation a footnote with
the web address of the project's home page must be provided.

\sectionb{3}{THE MODELS}

The spectrophotometric characteristics of the galaxy model are defined
by the geometry of stellar and dust mass distribution and their spectral
properties.  Stars and interstellar dust can be distributed in the model
volume according to a wide range of different 2-D laws:  single and
double exponential, spheroidal, de Vaucouleurs, etc.  Due to a
limitation, imposed by the used algorithm, all interstellar dust
distributed within the galaxy models has the same optical
characteristics:  albedo, scattering phase function and absorptivity
(Laor \& Draine 1993), while each stellar population, included in the
galaxy model, can possess its unique spectral energy distributions.

The model galaxy description is passed into the GFE program through the
input file {\tt galaxy}, which must contain one mandatory and any number
of optional blocks, see Figure~1.  The mandatory block defines a general
model geometry, parameters and execution control flags, which determine
computation steps to be performed.  This block must be located at the
beginning of the {\tt galaxy} file.  Optional blocks start with the line
containing a single integer denoting its type:  mass distribution of
stellar population and its spectrum (block type 1), interstellar dust
mass distribution (2) and dust grain optical parameters (3).  Optional
blocks may appear in the input file in arbitrary orders and in unlimited
numbers.  However, if the block 3 is present in the input file more than
once, the optical properties of dust grains will be computed using the
latest occurrence of this block, i.e., will override any previous
definitions.

The file {\tt galaxy} can also contain ``include'' and ``substitute''
directives and ``comments'', making it possible to combine an input file
from several files, each describing a particular stellar population
or dust distribution, etc. Included files may in turn contain further
comments and directives. It is also possible to sequentially compute
several models in a single GFE run by using ``substitute'' directives
and a substitute file, {\tt substab}. If during parsing of the main
input file {\tt galaxy} the program encounters symbol `{\tt \#}'
followed by a single digit, it attempts to substitute this pair of
symbols with a value from the {\tt substab} file. Each of the nine
possible substitute parameters (variables) from {\tt \#1} to {\tt \#9} 
corresponds to the first nine columns in the {\tt substab} file, each 
line of this file thus defining a set of variables for a single model. 
After finishing the computations the program returns to the start of 
the input file and repeats all steps, however, this time using 
the next line in the {\tt substab} file and proceeding until 
the end of that file.

At the output the program produces image files, aperture photometry
data and image cross-section (light-scans, integrated within one
pixel-wide image strips) data at each specified wavelength, as well as
an execution log file, containing information on the performed
computation steps, and model radiative energy balance after each
iteration.  The resulting aperture photometry is presented in magnitudes
by integrating output images within either circular or elliptical
(inclination dependent -- the aperture axis ratio is determined by the
stellar population scale-length to scale-height ratio) apertures of
specified sizes and as a photometric profile.

\begin{figure}
\centerline{\psfig{figure=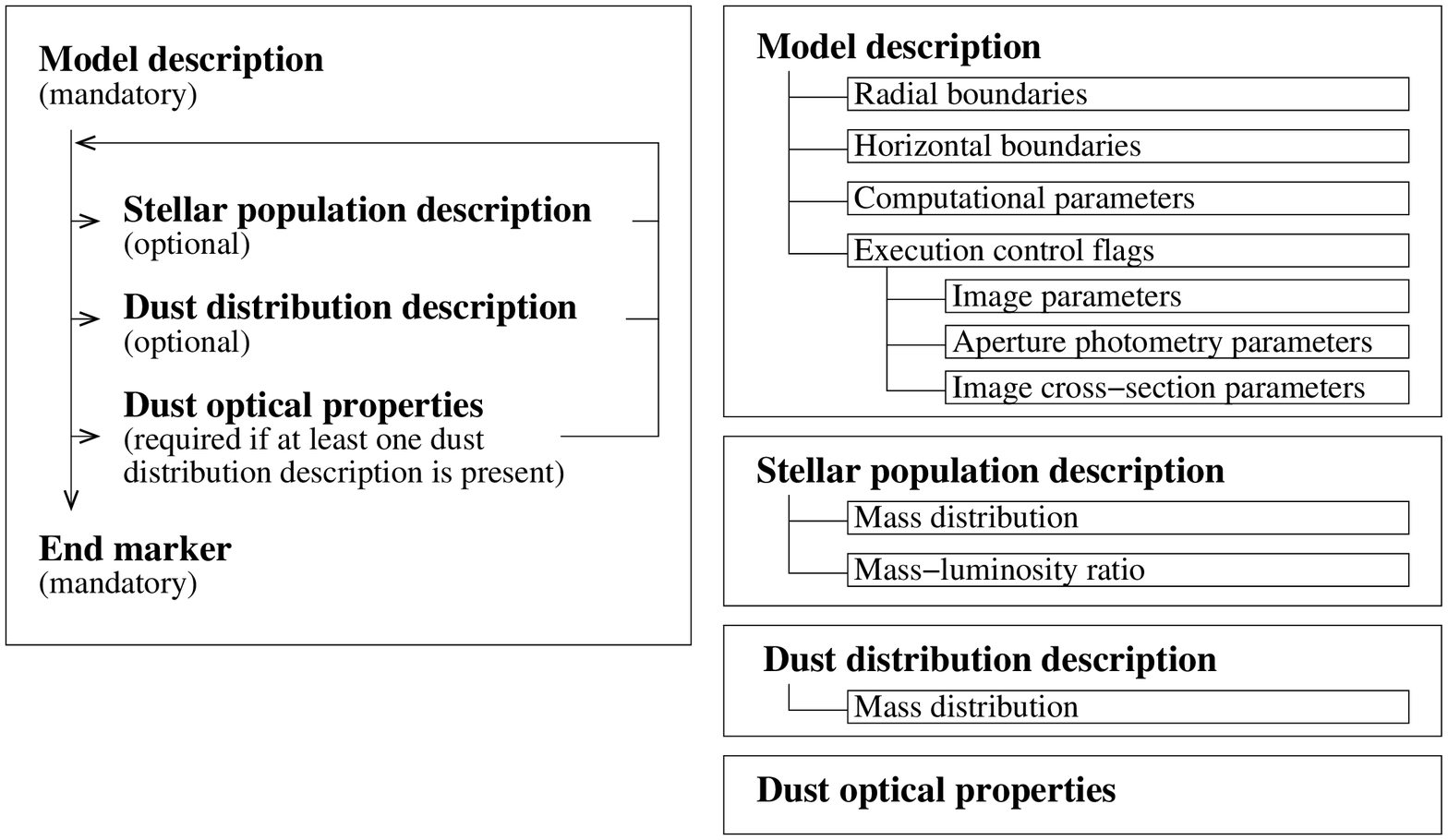,width=120mm,angle=0,clip=}}
\captionb{1} {The structure of the main model description file, {\tt
galaxy}, used in the GFE (left)  and the internal structure of each 
parameter block (right).}
\end{figure}

\sectionb{4}{SUMMARY}

The GFE program suite, presented in this paper, was successfully applied
to model a variety of dusty disk galaxies (Semionov \& Vansevi{\v c}ius
2002, 2003), even reproducing the effects of dusty spiral arms
(Semionov et al. 2006). By applying the GFE code it has been proved, that
the ray-tracing algorithm is a viable way to solve the radiative transfer
problem in astrophysics, which is an accurate enough, easily modifiable
and flexible tool for spectrophotometric modeling of galaxies.

\thanks{ We are thankful to Valdas Vansevi{\v c}ius for correcting the
manuscript.  This study was financially supported in part by a Grant
T-08/06 of the Lithuanian State Science and Studies Foundation.}

\vskip5mm

\References

\refb Baes M., Dejonghe H. 2001, MNRAS, 326, 722

\refb Ciardi B., Ferrara A., Marri S., Raimondo G. 2001, MNRAS, 324, 381

\refb Dullemond C., Turolla R. 2000, A\&A, 360, 1187

\refb Henyey L. 1937, ApJ, 85, 107

\refb Ivezi{\' c} {\v Z}., Groenewegen M., Men'shikov A., Szczerba R.
1997, MNRAS, 291, 121

\refb Laor A., Draine B. 1993, ApJ, 402, 441

\refb Pascucci I., Wolf S., Steinacker J., Dullemond C., Henning Th.,
Niccolini G., Woitke P., Lopez B. 2004, A\&A, 417, 793

\refb Razoumov A., Scott D. 1999, MNRAS, 309, 287

\refb Semionov D., Kodaira K., Stonkut{\. e} R., Vansevi{\v c}ius V.
2006, Baltic Astronomy, 15, 581 (this issue)

\refb Semionov D., Vansevi{\v c}ius V. 2002, Baltic Astronomy, 11, 537

\refb Semionov D., Vansevi{\v c}ius V. 2003, Baltic Astronomy, 12, 633

\refb Semionov D., Vansevi{\v c}ius V. 2005a, Baltic Astronomy, 14, 235

\refb Semionov D., Vansevi{\v c}ius V. 2005b, Baltic Astronomy, 14, 245

\refb Semionov D., Vansevi{\v c}ius V. 2005c, Baltic Astronomy, 14, 543

\end{document}